\documentclass[twocolumn,twoside,slac_two]{revtex4}
\usepackage{graphicx}
\usepackage{fancyhdr}
\begin{document}

\title{Four Years of {\em Fermi} LAT Observations of Narrow-Line Seyfert 1 Galaxies}


\author{F. D'Ammando, G. Tosti}
\affiliation{University of Perugia, Physics Department, and INFN, Via
  A. Pascoli, Perugia, Italy, I-06123}
\author{M. Orienti}
\affiliation{INAF-IRA Bologna, Via P. Gobetti 101, Bologna, Italy, I-40129}

\author{J. Finke}
\affiliation{U.S. Naval Research Laboratory, Code 7653, 4555 Overlook Ave. SW,
  Washington, USA, DC 20375-5352}

\author{on behalf of the {\em Fermi} Large Area Telescope Collaboration}

\begin{abstract}
Before the launch of the {\em Fermi} satellite only two classes of AGN were
known to generate relativistic jets and thus emit up to the $\gamma$-ray
energy range: blazars and radio galaxies, both hosted in giant elliptical
galaxies. The first two years of observations by the Large Area Telescope
(LAT) on board {\em Fermi} confirmed that these two are the most numerous classes of identified sources in the extragalactic
$\gamma$-ray sky, but the discovery of variable $\gamma$-ray emission from 5 radio-loud Narrow-Line Seyfert
1 galaxies (NLSy1s) revealed the presence of a possible emerging third class of AGN with relativistic jets.
Considering also that NLSy1s are typically hosted in spiral galaxy, this finding poses intriguing questions
about the nature of these objects, the onset of production of relativistic
jets, and the cosmological evolution of radio-loud AGN. Here, we report on a
preliminary investigation of the properties of this sample of radio-loud NLSy1
at MeV-GeV photon energies, utilizing the four-year accumulation of {\em Fermi} LAT
data. In addition we briefly discuss some radio-to-gamma-rays properties of the
$\gamma$-ray emitting NLSy1 in the context of the blazar scenario.

\end{abstract}

\maketitle

\thispagestyle{fancy}


\section{Introduction}

Only a small percentage of Active Galactic Nuclei (AGNs) are radio-loud, and
this characteristic is commonly ascribed to the presence of a relativistic jet, roughly perpendicular to the accretion disk. Accretion of gas onto the
super-massive black hole (SMBH) is
thought to power these collimated jets, even if the nature of the coupling
between the accretion disk and the jet is still among the outstanding open
questions in high-energy astrophysics \cite[e.g.][]{meier03}. Certainly relativistic jets are the most
extreme expression of the power than can be generated by a SMBH in the center
of an AGN. These objects have a total bolometric luminosity of up to 10$^{49-50}$ erg s$^{-1}$
\cite[e.g.][]{ackermann10, bonnoli11}, with a large fraction of
the power emitted in $\gamma$ rays. Before the launch of the {\it Fermi} satellite only two classes of AGNs were known
to generate these structures and thus to emit up to the $\gamma$-ray energy range: blazars and radio galaxies, both hosted in giant elliptical
galaxies \cite[]{blandford78}. The first 2 years of observation by
{\it Fermi}-LAT confirmed that the extragalactic $\gamma$-ray sky is dominated
by these two classes \cite[]{nolan12}. However, the discovery by the Large Area Telescope (LAT) on board the {\it Fermi}
satellite of variable $\gamma$-ray emission from a few radio-loud Narrow-Line Seyfert 1s (NLSy1s) revealed the presence of a
possible third class of AGNs with relativistic jets
\cite[]{abdo2009a,abdo2009b,abdo2009c}. On the contrary, no radio-quiet
Seyfert galaxies were detected in $\gamma$ rays until now
\cite[]{ackermann12b}. This finding poses intriguing questions
about the knowledge of the development of relativistic jets, the disk/jet connection, the Unification
model for AGNs and the evolution of radio-loud AGNs.

Even if the physics necessary to explain the generation, collimation and
evolution of the jet is still to be known \cite[e.g.][]{blandford01,blandford08}, there
is increasing evidence that the properties of jets in AGNs could be related to the
properties of the accretion flow that feeds the central SMBH. Recently,
\cite{ghisellini08,ghisellini11}  suggested a transition between BL Lac
objects and Flat Spectrum Radio Quasars (FSRQs) that can be mainly justified by
the different accretion regimes: sub-Eddington in the first class of objects, leading to radiatively
inefficient accretion flows and relatively weak jets, whereas near-Eddington
in the second class, giving rise to bright disks and powerful jets. This
scenario seems to be also in agreement with the separation of their ``non-beamed'' counterparts, Fanaroff-Riley I
and Fanaroff-Riley II radio galaxies. In this context, the case of the
peculiar radio-loud NLSy1s has received increasing attention. 

NLSy1 is a class of AGN identified by \cite{osterbrock85} and characterized by their optical properties: narrow permitted lines (FWHM
(H$\beta$) $<$ 2000 km s$^{-1}$) emitted from the broad line region,
[OIII]/H$\beta$ $<$ 3, and a FeII bump \cite[for a review see e.g.][]{pogge00}. They also exhibit strong X-ray variability, steep X-ray spectra,
substantial soft X-ray excess and relatively high luminosity. These characteristics point to systems with smaller masses of the central BH (10$^6$-10$^8$ M$_\odot$) and higher accretion rates
(close to or above the Eddington limit) with respect to blazars and radio
galaxies. NLSy1 are generally radio-quiet, with only a small fraction of them
\cite[$<$ 7$\%$,][]{komossa06} classified as radio-loud, and objects with high
values of radio-loudness (R $>$ 100) are even more sparse ($\sim$2.5\%), while $\sim$15$\%$
of quasars are radio-loud. In the past, several authors inferred from studies
of non-simultaneous radio to X-ray data different possibilities on the nature
of these objects. \cite{komossa06} argued that radio-loud NLSy1s could be some
young stage of quasars, while \cite{yuan08} claimed that some NLSy1 might be high-frequency-peaked flat spectrum radio quasars conjectured by
\cite{padovani07} and not predicted by the blazar sequence, but no conclusive
evidence has been presented. 
However, as shown with preliminary results in \cite{foschini09}, some radio-loud NLSy1s show similarities with FSRQs, while others are like BL
 Lacs. Moreover, as in the case of blazars and radio galaxies, there should be
 a ``parent population'' with the jet viewed at large angles. The first
 source of this type could be PKS 0558$-$504 \cite[]{gliozzi10}. Other
 candidates have been detected by \cite{doi12}. Therefore, it
 is possible that NLSy1s are a set of
 low-mass systems parallel to blazars and radio galaxies.

The firm confirmation of the existence of relativistic jets also in
  Seyfert galaxies, provided by the LAT detection of NLSy1s, opened a large and
  unexplored research space for important discoveries for our knowledge of the
  AGNs, but brought with itself new challenging questions. What are the
  differences between this class of $\gamma$-ray AGNs and blazars and radio galaxies? What is the origin of the radio-loudness? What
  are the parameters determining the jet formation and how it is possible to
  have the formation of a jet in a population of AGN mostly radio-quiet such as the
  NLSy1s? Is there a limiting BH mass above which objects are
  preferentially radio-loud? How do these objects
  fit into the blazar sequence? Four years
after the announcement of the detection of the first $\gamma$-ray NLSy1 by
{\it Fermi}-LAT, PMN J0948$+$0022 \cite[]{abdo2009a}, only a few indications about the nature of these objects have been obtained.
In Section 2 we report the LAT data analysis and results of the 5 $\gamma$-ray
NLSy1s over 4 years of {\em Fermi} observations, focusing on the two flaring
sources SBS 0846$+$513 and PMN J0948$+$0022 in Section 3. In Section 4 and 5
we discuss the host galaxies and the jet formation, respectively, for the
NLSy1s. In Section 6 we report about searching new $\gamma$-ray NLSy1s, while
concluding remarks are presented in Section 7.

\section{The $\gamma$-ray view of NLSy1}

Up to now 5 radio-loud NLSy1 galaxies have been detected at high significance by {\em Fermi}-LAT: PMN J0948$+$0022, PKS 1502$+$036, 1H 0323$+$342, PKS
2004$-$447, and SBS 0846$+$513. Here we analyze the first four years of
$\gamma$-ray observations of these sources.
The LAT data reported in this paper were collected from 2008 August 4 (MJD
54682) to 2012 August 4 (MJD 56143). During this time the LAT instrument
operated almost entirely in survey mode. The analysis was performed with the \texttt{ScienceTools} software package version v9r27p1. The LAT data
were extracted within a $10^{\circ}$ Region of Interest (RoI) centred at the
location of the 5 NLSy1s. Only events belonging to the ``Source'' class were used. In addition, a cut on the
zenith angle ($< 100^{\circ}$) was also applied to reduce contamination from
the Earth limb $\gamma$ rays, which are produced by cosmic rays interacting with the upper atmosphere. 
The spectral analysis was performed with the instrument response functions (IRFs)
\texttt{P7SOURCE\_V6} using an unbinned maximum likelihood method implemented
in the Science tool \texttt{gtlike}. A Galactic diffuse emission model and isotropic component, which is the sum of
an extragalactic and instrumental background were used to model the
background\footnote{http://fermi.gsfc.nasa.gov/ssc/data/access/lat/Background\\Models.html}. The
normalizations of both components in the background model were allowed to vary
freely during the spectral fitting. 

We evaluated the significance of the $\gamma$-ray signal from the sources by
means of the Test Statistics TS = 2$\Delta$log(likelihood) between models with
and without the source \cite[]{mattox96}. The source model used in
\texttt{gtlike} includes all the point sources from the 2FGL catalogue that
fall within $20^{\circ}$ from the target source. The spectra of these sources
were parametrized by power-law functions, $dN/dE \propto$
$(E/E_{0})^{-\Gamma}$, where $\Gamma$ is the photon index, or log-parabola,
$dN/dE \propto$ $(E/E_{0})^{-\alpha-\beta \, \log(E/E_0)}$, where $E_{0}$ is a
reference energy, $\alpha$ the spectral slope at the energy $E_{0}$, and the parameter $\beta$ measures the
curvature around the peak. 
A first maximum likelihood was performed to remove from the model the sources
having TS $<$ 10 and/or the predicted number of counts based on the fitted model $N_{pred} < 3 $. A second maximum likelihood was performed
on the updated source model. The fitting procedure has been performed with the
sources within 10$^{\circ}$ from the target source included with the
normalization factors and the photon indices left as free parameters. For the
sources located between 10$^{\circ}$ and 20$^{\circ}$ from our target we kept the
normalization and the photon index fixed to the values of the 2FGL
catalog. We used a power-law model for PKS 1502$+$036 and PKS 2004$-$447, and a
log-parabola for 1H 0323$+$342 and PMN J0948$+$0022, as in the 2FGL
catalog. For SBS 0846$+$513 we used a power-law model as in \cite{dammando12}
for the third year of {\em Fermi} observation. 

The results of the LAT analysis over the entire period are summarized
in Table~\ref{LAT}. The average observed isotropic luminosity of the 5 objects
in the 0.1--100 GeV energy range spans between 10$^{44}$ erg s$^{-1}$ and 10$^{47}$ erg
s$^{-1}$, a range of values typical of blazars. This could be an indication of
a similar viewing angle with respect to the jet axis and beaming factor for
the $\gamma$-ray emission between blazars and $\gamma$-ray NLSy1s. In
particular, SBS 0846$+$513 and PMN J0948$+$0022 showed $\gamma$-ray
flaring activity combined with a moderate spectral evolution \cite[]{dammando12,foschini11}, a behaviour already
observed in bright FSRQs and low-synchrotron-peaked (LSP) BL Lacs detected by {\em
  Fermi} LAT \cite[]{abdo10}. The average photon index of the 5 NLSy1s spans a
range ($\Gamma$ = 2.2--2-6) narrower than those observed in Misaligned AGNs
\cite[$\Gamma$ = 1.9--2.7,][]{grandi12} and similar to the average values
observed for FSRQs and LSP BL Lacs \cite[]{nolan12}.

\begin{table*}
\caption{Gamma-ray characteristics of radio-loud NLSy1 galaxies detected by
  {\em Fermi}-LAT.}
\begin{center}
\begin{tabular}{llccccc}
\hline \hline
\multicolumn{1}{c}{Source} &
\multicolumn{1}{c}{Redshift} &
\multicolumn{1}{c}{Flux (E $>$ 100 MeV)} &
\multicolumn{1}{c}{Photon index} &
\multicolumn{1}{c}{Curvature} &
\multicolumn{1}{c}{TS}  &
\multicolumn{1}{c}{L$_{\gamma}$}  \\
\multicolumn{1}{c}{} &
\multicolumn{1}{c}{z} &
\multicolumn{1}{c}{$\times$10$^{-8}$ ph cm$^{-2}$ s$^{-1}$} &
\multicolumn{1}{c}{$\Gamma$/$\alpha$} &
\multicolumn{1}{c}{$\beta$} &
\multicolumn{1}{c}{}  &
\multicolumn{1}{c}{$\times$10$^{46}$erg s$^{-1}$} 
\\
\hline
1H 0323$+$342 & 0.061 & 3.52$\pm$0.42 & 2.62$\pm$0.13 & 0.63$\pm$0.17 & 199 & 0.01 \\
SBS 0846$+$513 & 0.5835 & 2.63$\pm$0.22   & 2.18$\pm$0.05  & -  & 658 & 2.95 \\
PMN J0948$+$0022 & 0.585 & 11.32$\pm$0.43 & 2.37$\pm$0.05  &  0.23$\pm$0.03 & 2287 & 10.02 \\
PKS 1502$+$036 & 0.409 & 4.14$\pm$0.39  &  2.60$\pm$0.06 & - & 309 & 1.25 \\
PKS 2004$-$447 & 0.24 & 1.37$\pm$0.33   & 2.56$\pm$0.14 & - & 49 & 0.11 \\
\hline
\hline
\end{tabular}
\end{center}
\label{LAT}
\end{table*}

\begin{figure}[t]
\centering
\includegraphics[width=85mm]{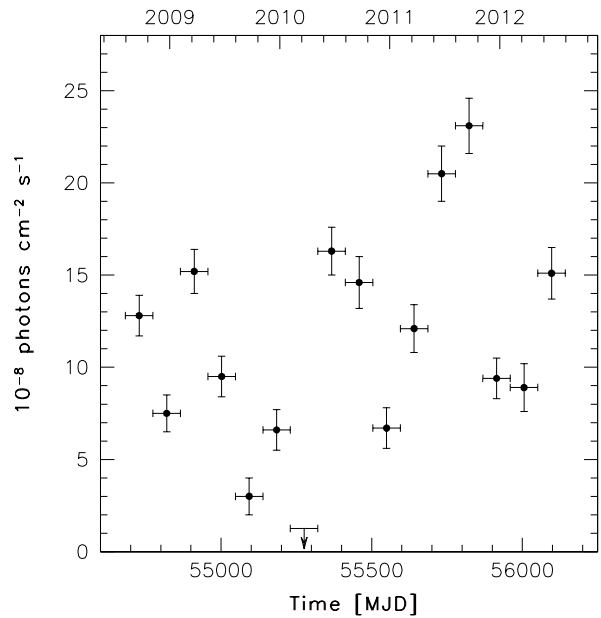}
\caption{Integrated flux light curve of PMN J0948$+$0022 obtained in the 0.1--100 GeV energy range during 2008 August 4 -- 2012 August 4 with 3-month time bins. Arrows refer to 2-$\sigma$ upper limits on the source flux. Upper limits are computed when TS $<$ 10.} \label{0948}
\end{figure}

\begin{figure}[t]
\centering
\includegraphics[width=85mm]{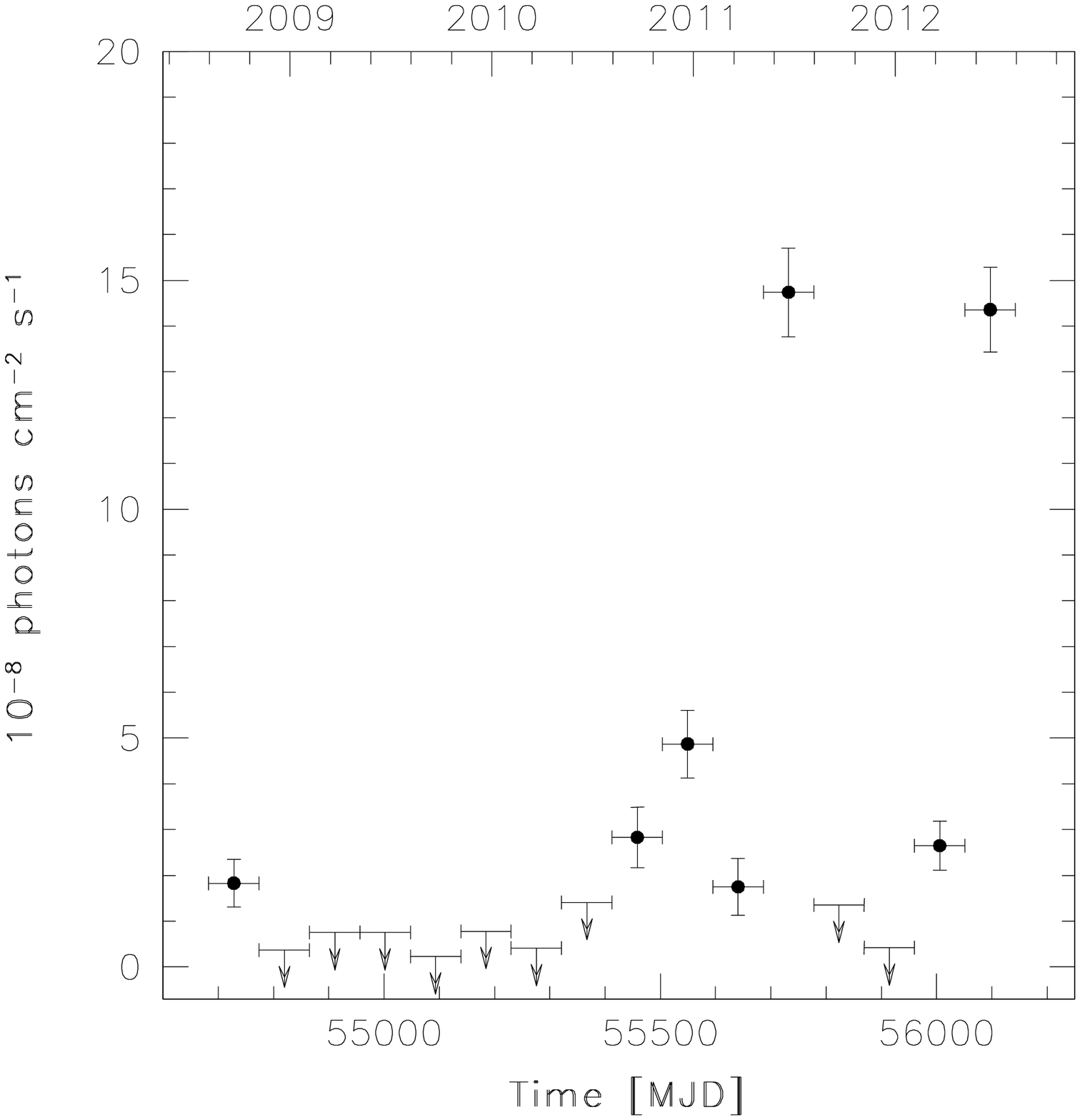}
\caption{Integrated flux light curve of SBS 0846$+$513 obtained in the 0.1--100 GeV energy range during 2008 August 4 -- 2012 August 4 with 3-month time bins. Arrows refer to 2-$\sigma$ upper limits on the source flux. Upper limits are computed when TS $<$ 10.} \label{0846}
\end{figure}

\begin{figure}[t]
\centering
\includegraphics[width=85mm]{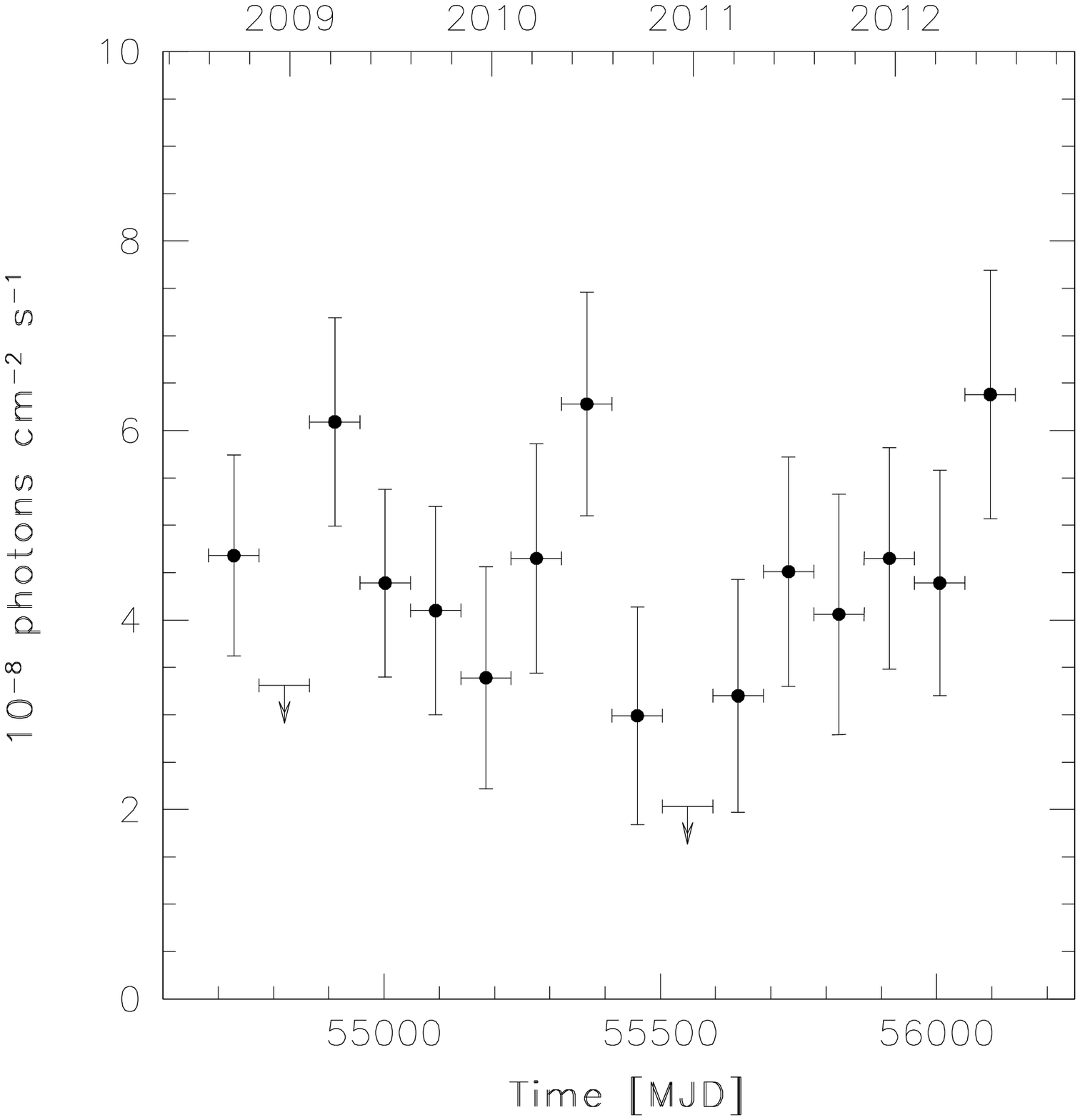}
\caption{Integrated flux light curve of PKS 1502$+$036 obtained in the 0.1--100 GeV energy range during 2008 August 4 -- 2012 August 4 with 3-month time bins. Arrows refer to 2-$\sigma$ upper limits on the source flux. Upper limits are computed when TS $<$ 10.} \label{1502}
\end{figure}

\begin{figure}[t]
\centering
\includegraphics[width=85mm]{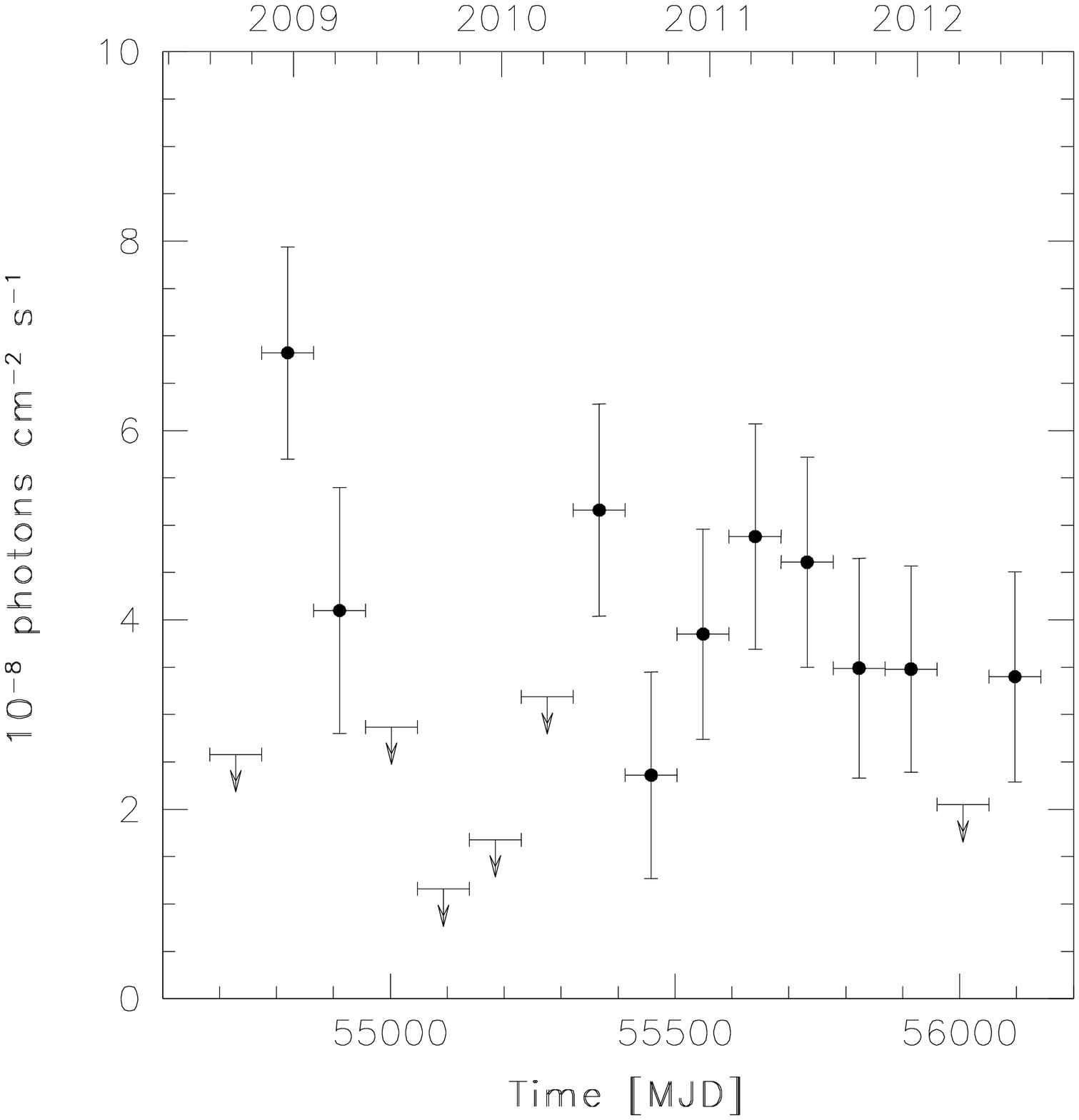}
\caption{Integrated flux light curve of 1H 0323$+$342 obtained in the 0.1--100 GeV energy range during 2008 August 4 -- 2012 August 4 with 3-month time bins. Arrows refer to 2-$\sigma$ upper limits on the source flux. Upper limits are computed when TS $<$ 10.} \label{0323}
\end{figure}

\begin{figure}[t]
\centering
\includegraphics[width=85mm]{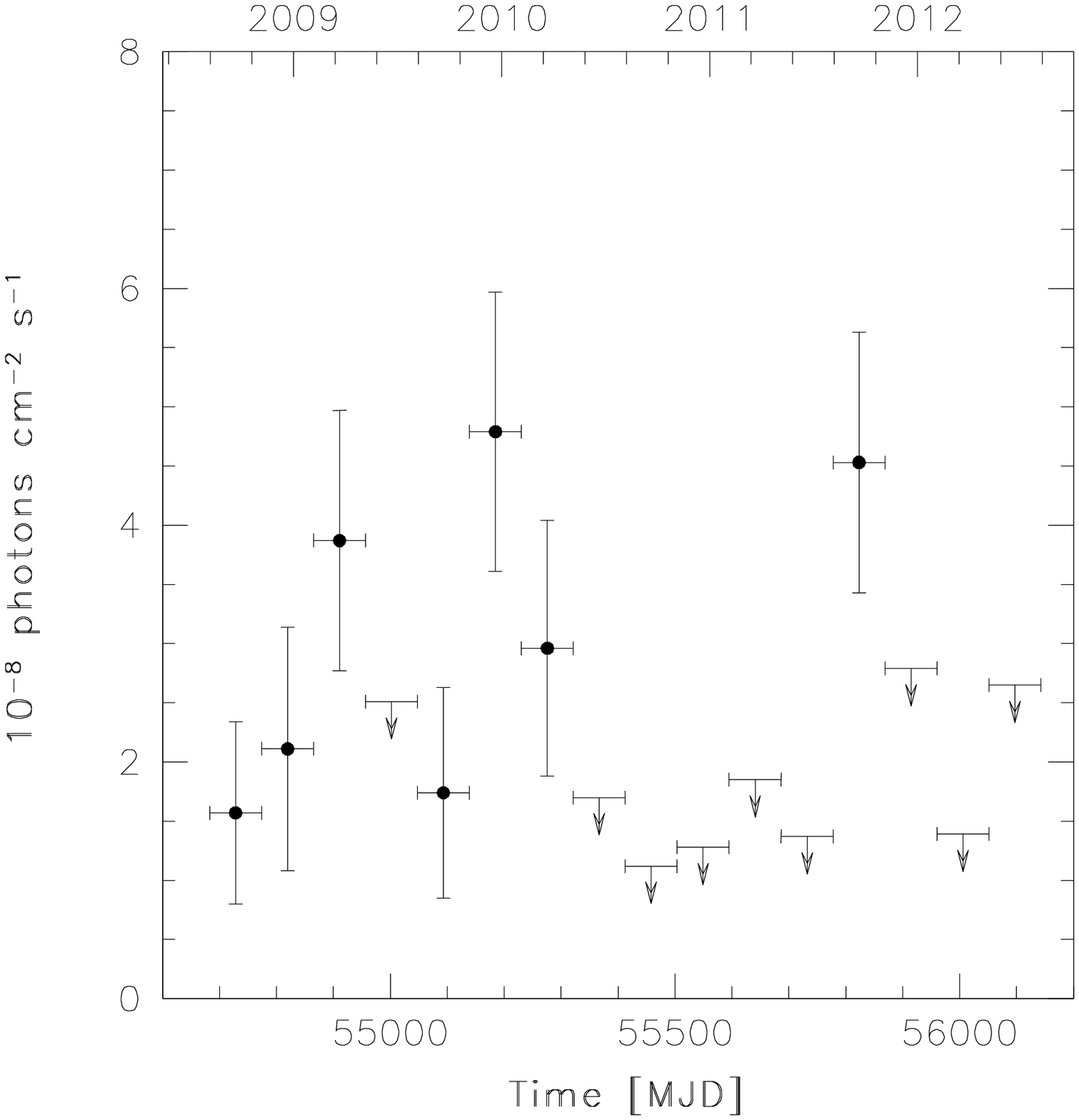}
\caption{Integrated flux light curve of PKS 2004$-$447 obtained in the 0.1--100 GeV energy range during 2008 August 4 -- 2012 August 4 with 3-month time bins. Arrows refer to 2-$\sigma$ upper limits on the source flux. Upper limits are computed when TS $<$ 10.} \label{2004}
\end{figure}

Figures~\ref{0948} to \ref{2004} show the $\gamma$-ray light curves of the 5 NLSy1s for the period 2008
August 4 --2012 August 4 using 3-month time bins. For each time bin the spectral parameters
of the target source and all sources within 10$^{\circ}$ from it were frozen to the value resulting
from the likelihood analysis over the entire period. If TS $<$ 10 the value of
the fluxes were replaced by the 2-$\sigma$ upper limits. The systematic
uncertainty in the flux is energy dependent: it amounts to $10\%$ at 100 MeV, decreasing to
$5\%$ at 560 MeV, and increasing again to $10\%$ above 10 GeV
\cite[]{ackermann12}. A different level of activity has been clearly observed in
these five objects, with significant variability in SBS 0846$+$513 and PMN J0948$+$0022.

\section{Flaring NLSy1s: PMN J0948$+$0022 and SBS 0846$+$513}

One of the key question is the maximum power released by the jets of
radio-loud NLSy1, and for this reason $\gamma$-ray flaring activities from
these sources are catalyzing a growing interest in the astrophysical
community. The first answers arrived in 2010 July when PMN J0948$+$0022 underwent a $\gamma$-ray flaring activity
with a daily peak flux of $\sim$1$\times$10$^{-6}$ ph cm$^{-2}$ s$^{-1}$
\citep{foschini11}. The first spectral energy distributions (SEDs) collected
for the 4 $\gamma$-ray NLSy1s detected in the first year of {\em Fermi} operation showed clear similarities with blazars: a double-humped
shape with disk component in UV, physical parameters blazar-like and jet power
in the average range of blazars \cite[]{abdo2009c}.
The comparison of the SED of PMN J0948$+$0022 during the 2010 July flaring
activity with that of 3C 273, a typical FSRQ, shows a more extreme Compton
dominance in the NLSy1. The disagreement of the two SEDs can be due to the difference in BH masses and Doppler factor of the two jets. 
Further $\gamma$-ray flaring episodes from PMN J0948$+$0022 have been observed
in 2011 June \cite[]{dammando11} and 2013 January \cite[]{dammando13}. 

A strong $\gamma$-ray flare was observed also
from SBS 0846$+$513 in 2011 June, reaching an isotropic $\gamma$-ray
luminosity (0.1--300 GeV) of $\sim$10$^{48}$ erg s$^{-1}$, comparable to that of the bright FSRQs \citep{dammando12}. Variability and
spectral properties of SBS 0846$+$513 in radio and $\gamma$ rays indicate a blazar-like
behaviour. In addition, from the model-fitting of 4-epoch MOJAVE data of this
source in 2010-2012 we found
that the jet components are separating with an apparent velocity of
(10.9$\pm$1.4)$c$. This value suggest the presence of boosting effect as well as
in blazars \cite[]{dammando12b}. 
The power released
during the flaring activity and the apparent superluminal motion velocity are
strong indications of the presence of a relativistic jets as powerful as those in blazars, despite the lower BH masses.

\section{Host galaxy}

The discovery of a relativistic jet in a class of AGN usually hosted in spiral
galaxies such as the NLSy1s was a great surprise. Unfortunately only very
sparse observations of the host galaxy of radio-loud NLSy1s are available up to
now and the sample of objects studied by \cite{deo06} and \cite{zhou06} had z
$<$ 0.03 and z $<$ 0.1, respectively, including both radio-quiet and
radio-loud sources.

Among the $\gamma$-ray emitting NLSy1s detected by {\em Fermi} LAT only for 1H
0323$+$342 {\em Hubble Space Telescope} (HST) and Nordic Optical Telescope observations
are available. These observations revealed a one-armed galaxy morphology or a
circumnuclear ring, suggesting two possibilities: the spiral arms of the host
galaxy \cite[]{zhou07} or the residual of a galaxy merging \cite[]{anton08}. On
the other hand, no significant resolved structures have been observed instead by HST for SBS
0846$+$513 \cite[]{maoz93}, and no high-resolution observations are available for the
remaining $\gamma$-ray NLSy1s. Thus the possibility that the development of
relativistic jets in these objects could be due to strong merger activity, not
so usual in disk/spiral galaxies, cannot be ruled out. Further high-resolution 
observations of the host galaxy of $\gamma$-ray NLSy1s will be fundamental to obtain
important insights into the onset of production of relativistic jets.

\section{Radio-loudness and jet formation}

The mechanism at work for producing a relativistic jet is not still clear. In particular the physical parameters that
drive the jet formation is still under debate. One fundamental parameter could
be the BH mass, with only large masses that can allow efficient formation
of a relativistic jet. \cite{sikora07} suggested that AGN with M$_{\rm BH}$ $>$ 10$^{8}$ M$_\odot$ have a radio loudness 3 orders of
magnitude greater than the AGN with M$_{\rm BH}$ $<$ 3 $\times$ 10$^{7}$
M$_\odot$. According to the ``modified spin paradigm'' proposed, another fundamental
parameter for the efficiency of a relativistic jet production should be the BH
spin, with SMBHs in elliptical galaxies having on average much larger spins than SMBHs in spiral galaxies. This is due
to the fact that the spiral galaxies are characterized by multiple accretion
events with random angular momentum orientation and small increments
of mass, while elliptical galaxies underwent at least one major merger with
large matter accretion triggering an efficient spin-up of the SMBHs. 
The accretion rate (thus the mass) and the spin of the BH seem to
be related to the host galaxy, leading to the hypothesis that relativistic jets
can develop only in elliptical galaxy \cite[e.g.][]{marscher09,bottcher02}. In this
context the large radio-loudness of SBS 0846$+$513 could challenge this idea
if the BH mass (8.2$\times$10$^{6}$-5.2$\times$10$^{7}$ M$_{odot}$) estimated
by \cite{zhou05} is confirmed.

We noted that BH masses of radio-loud NLSy1s are generally larger with respect
to the entire sample of NLSy1s \cite[M$_{\rm BH}$ $\approx$(2--10) $\times$10$^{7}$ M$_\odot$;][]{komossa06,yuan08}, even if still small when
compared to radio-loud quasars. The larger BH masses of radio-loud NLSy1s
could be related to prolonged accretion episodes that can spin-up the BHs. The
small fraction of radio-loud NLSy1s with respect to radio-loud quasars could be
an indication that not in all of the formers the high-accretion regime lasted
long enough to spin-up the central BH \cite[]{sikora09}.

\section{New $\gamma$-ray emitting NLSy1?}

In addition to the 5 NLSy1s already observed new $\gamma$-ray emitting
NLSy1 could be detected accumulating more and more {\em Fermi}-LAT data. For
this reason we started from a list of 39 NLSy1s reported in \cite{yuan08,
  zhou02, whalen06, komossa06, oshlack01} with a radio-loudness $R > 20$ and
analyzed 4 years of LAT data to search systematically for GeV emission from them. No significant (TS $>$ 25) detection of new $\gamma$-ray NLSy1s
over the entire 4-year period was obtained. 

Anyway the variability could be a key ingredient to take into account for a
high-significance detection of these sources, as clearly showed by SBS
0846$+$513 that after two years of very low $\gamma$-ray activity \cite[and for this reason not included
in the First or Second {\em Fermi}-LAT catalog,][]{abdo10,nolan12} started an increase of the flux up to
the flaring activity detected in 2011 June--July \cite[]{dammando12}. For this
reason the next step will be an investigation of the activity of these
candidates $\gamma$-ray NLSy1 on different time scales (e.g.~1 month, 3 months, 1 year).

\section{Concluding remarks}

The presence of a relativistic jet in some radio-loud NLSy1 galaxies, first
suggested by their variable radio emission and flat radio spectra, is now
confirmed by the {\em Fermi}-LAT detection of five NLSy1s
in $\gamma$ rays. The flaring episodes observed in $\gamma$ rays from SBS 0846$+$513 and PMN
J0948$+$0022 are strong indications of the presence of a relativistic jet as
powerful as those of blazars. These two sources showed all the characteristics 
of the blazar phenomenon and they could be a relatively low mass (and possibly younger) version of blazars. Further multifrequency observations of
these and the other $\gamma$-ray emitting NLSy1s will be fundamental for
investigating in detail their characteristics over the entire electromagnetic
spectrum. The impact of the properties of the central engines in radio-loud
NLSy1s, which seem quite different of those of quasar and manifest in their
peculiar optical characteristics, on the $\gamma$-ray emission mechanisms is
currently under debate. In addition, the detection of relativistic jets in a class of AGN usually hosted in spiral galaxies is very
intriguing, challenging the theoretical scenario of relativistic jet formation
proposed up to now. The detection of new NLSy1s in $\gamma$ rays by {\em Fermi}-LAT will be important for extending the sample and better
characterizing this new class of $\gamma$-ray emitting AGN.

\begin{acknowledgments}
The {\em Fermi} LAT Collaboration acknowledges generous ongoing
support from a number of agencies and institutes that have supported
both the development and the operation of the LAT as well as
scientific data analysis.  These include the National Aeronautics and
Space Administration and the Department of Energy in the United
States, the Commissariat \`a l'Energie Atomique and the Centre
National de la Recherche Scientifique / Institut National de Physique
Nucl\'eaire et de Physique des Particules in France, the Agenzia
Spaziale Italiana and the Istituto Nazionale di Fisica Nucleare in
Italy, the Ministry of Education, Culture, Sports, Science and
Technology (MEXT), High Energy Accelerator Research Organization (KEK)
and Japan Aerospace Exploration Agency (JAXA) in Japan, and the
K.~A.~Wallenberg Foundation, the Swedish Research Council and the
Swedish National Space Board in Sweden. Additional support for science
analysis during the operations phase is gratefully acknowledged from
the Istituto Nazionale di Astrofisica in Italy and the Centre National
d'\'Etudes Spatiales in France. FD, MO acknowledge financial contribution from
grant PRIN-INAF-2011.
\end{acknowledgments}
 
\bigskip % extra skip inserted

\end{document}